\documentclass[12pt,a4paper,onecolumn,groupedaddress,preprintnumbers,nofootinbib,floatfix,prd]{revtex4}
\usepackage[top=3cm, bottom=2.5cm, left=2.5cm, right=2.5cm]{geometry}
\pdfoutput=1

\usepackage[utf8]{inputenc}
\usepackage[T1]{fontenc}
\usepackage{commath}
\usepackage{graphicx}
\usepackage{amsmath,amsthm,amssymb,amsfonts}
\usepackage{color}
\usepackage{subfigure}
\usepackage{natbib}
\usepackage{hyperref}
\hypersetup{
    bookmarks=true,         
    unicode=false,          
    pdftoolbar=true,        
    pdfmenubar=true,        
    pdffitwindow=false,     
    pdfstartview={FitH},    
    pdftitle={My title},    
    pdfauthor={Author},     
    pdfsubject={Subject},   
    pdfcreator={Creator},   
    pdfproducer={Producer}, 
    pdfkeywords={keyword1} {key2} {key3}, 
    pdfnewwindow=true,      
    colorlinks=true,       
    linkcolor=magenta,          
    citecolor=magenta,        
    filecolor=magenta,      
    urlcolor=cyan           
}

\newcommand{\be}{\begin{equation}}
\newcommand{\ee}{\end{equation}}
\newcommand{\beq}{\begin{eqnarray}}
\newcommand{\eeq}{\end{eqnarray}}

\newcommand{\bmp}{\noindent\begin{minipage}{16cm}}
\newcommand{\emp}{\end{minipage}\vskip 7mm} 

\begin{document}
\title{Quasinormal modes and thermalization in \\ Improved Holographic QCD}

\author{Timo Alho}
\email[]{alho@hi.is}
\affiliation{Helsinki Institute of Physics and Department of Physics, University of Helsinki} 

\author{Jere Remes}
\email[]{jere.remes@helsinki.fi}
\affiliation{Helsinki Institute of Physics and Department of Physics, University of Helsinki}

\author{Kimmo Tuominen}
\email[]{kimmo.i.tuominen@helsinki.fi}
\affiliation{Helsinki Institute of Physics and Department of Physics, University of Helsinki}

\author{Aleksi Vuorinen}
\email[]{aleksi.vuorinen@helsinki.fi}
\affiliation{Helsinki Institute of Physics and Department of Physics, University of Helsinki}


\begin{abstract}
\noindent  Following a series of similar calculations in simpler non-conformal holographic setups, we determine the quasinormal mode spectrum for an operator dual to a gauge-invariant scalar field within the Improved Holographic QCD framework. At temperatures somewhat above the critical temperature of the deconfinement transition, we find a small number of clearly separated modes followed by a branch-cut-like structure parallel to the real axis, the presence of which is linked to the form of the IHQCD potential employed. The temperature dependence of the lowest nonzero mode is furthermore used to study the thermalization time of the corresponding correlator, which is found to be of the order of the inverse critical temperature near the phase transition and decrease slightly faster than $1/T$ at higher temperatures.
\end{abstract}

\preprint{HIP-2020-04/TH}

\maketitle

\section{Introduction}
\label{sec:intro}

The quantitative description of equilibration dynamics in gauge theories is a notoriously complicated problem, the study of which is motivated by the desire to understand the time evolution of systems such as ultrarelativistic heavy ion collisions or the expanding early universe \cite{Brambilla:2014jmp}. In the former context, where the dynamics is governed by the strong nuclear force, it becomes essential to describe thermalization away from the tractable weak coupling regime \cite{Kurkela:2011ti}. For this reason, the gauge/gravity duality has become a standard tool in the field \cite{CasalderreySolana:2011us}, and a considerable amount of work has indeed been devoted to studing equilibration in strongly coupled ${\mathcal N}=4$ Super Yang-Mills (SYM) theory by mapping the process to the gravitational dynamics of black hole (BH) formation in AdS spacetime \cite{Chesler:2008hg}. Important milestones in this line of research include e.g.~the successful description of shock wave collisions, the observation of a rapid onset of hydrodynamic behavior, and the subsequent understanding that this does not require a full thermalization or even isotropization of the system (see e.g.~\cite{Chesler:2010bi,Heller:2012km,Bantilan:2014sra,Waeber:2019nqd} and references therein).

While a considerable amount of physical insight has been gained from studies of the SYM theory in its infinitely strongly coupled regime, it is clear that the application of the results to real QCD requires an understanding of the quantitative effects of conformal invariance breaking as well as the finiteness of the gauge coupling \cite{Attems:2016tby,Grozdanov:2016vgg,Solana:2018pbk}. These issues have indeed been studied in many different contexts, in some cases involving even shock wave collisions \cite{Grozdanov:2016zjj,Attems:2016ugt,Attems:2017zam}, but often by restricting to the study of the quasinormal mode (QNM) spectra of BH solutions in different spacetimes \cite{Steineder:2012si,Waeber:2015oka,Waeber:2018dae}. This is understandable, since the QNM spectrum is a key quantity determining a holographic system's equilibration from small perturbations, as it is dual to the pole locations of the corresponding retarded Green's function on the field theory side. In fact, as argued in \cite{Keranen:2015mqc}, the thermalization time of a Wightman function in the field theory dual of a five-dimensional system undergoing gravitational collapse should be inversely proportional to the imaginary part of the lowest QNM determined for the same correlator in thermal equilibrium.

In the paper at hand, we continue work towards understanding equilibration in a QCD-like plasma, but staying on the level of QNMs. In particular, we study the retarded Green's function of an operator dual to a gauge-invariant scalar field in Improved Holographic QCD (IHQCD) ---  a five-dimensional bottom-up model constructed to mimic large-$N_c$ non-supersymmetric Yang-Mills (YM) theory \cite{Gursoy:2007cb,Gursoy:2007er,Gursoy:2009jd,Jarvinen:2011qe,Alho:2012mh,Alho:2013hsa,Alho:2015zua}. This work can be considered a direct continuation of similar studies of QNMs in related but often simplified non-conformal models \cite{Alanen:2011hh,Janik:2016btb,Betzios:2017dol,Betzios:2018kwn} (see also \cite{Gursoy:2013zxa,Ishii:2015gia,Janik:2015iry,Demircik:2016nhr,Janik:2017ykj,Critelli:2017oub}), differing from them through the use of the more realistic IHQCD potential introduced in \cite{Alho:2015zua}. Utilizing the fact that the logarithmic running of the QCD gauge coupling is built into IHQCD, we are able to study the QNM spectrum --- and thus the thermalization time scale --- away from the infinitely strongly coupled regime. As discussed in detail below, our results point towards the emergence of a branch-cut-like structure on the complex frequency plane in the limits of increased temperature and QNM mode number, which both probe the more weakly coupled UV limit of the theory.

Our paper is organized as follows. In Section \ref{sec:model}, we briefly review the model we work in as well as our strategy for determining its bulk thermodynamic properties as well as QNM spectra. The results of this investigation are then presented in Section \ref{sec:results}, while Section \ref{sec:checkout} is  devoted to a detailed discussion of our findings. Some computational details useful in the determination of the QNM spectra are finally relegated to Appendix \ref{sec:app}.

\section{Holographic model}
\label{sec:model}

In this section, we discuss first the construction and basic properties of the holographic model we work with, IHQCD, and after this present some details concerning the determination of different thermodynamic quantities as well as QNM spectra within this framework.

\subsection{Basic equations}

Improved holographic QCD is defined by the five-dimensional gravity action
\be
S=\frac{1}{16\pi G_5}\int d^5x\sqrt{-g}[R-\frac{4}{3}(\partial_\mu\phi)^2+V(\phi)],
\ee
together with the metric ansatz $ds^2=b(z)^2(-f(z)dt^2+d{\bf{x}}^2+{f(z)}^{-1}dz^2)$  \cite{Gursoy:2007cb, Gursoy:2007er}, where $z$ is a radial coordinate chosen in such a way that the UV boundary resides at $z=0$. Here, the scalar field $\phi(z)$ and the functions $b(z)$ and $f(z)$ are in turn determined from Einstein equations, which in this case take the forms
\begin{align}
	6\frac{\dot{b}^2}{b^2}+3\frac{\ddot{b}}{b}+3\frac{\dot{b}\dot{f}}{bf}&= \frac{b^2}{f}V(\phi), \\
	6\frac{\dot{b}^2}{b^2}-3\frac{\ddot{b}}{b} &= \frac{4}{3}\dot{\phi}^2, \\
	\frac{\ddot{f}}{\dot{f}}+3\frac{\dot{b}}{b} &=0,
	\label{Eq:Einstein}
\end{align}
with the dot denoting a derivative with respect to $z$.
The scalar field equation derived from Eqs. (2)--\eqref{Eq:Einstein} finally reads
\be \ddot\phi+\frac{d}{dz}\ln(fb^3)\dot\phi+\frac{3}{8}\frac{b^2}{f}V'(\phi)=0
\ee

In addition, both the $z$-coordinate and the function $b(z)$ are closely related to the renormalization scale on the field theory side, so that we may write
\be
\beta(\lambda)=b\frac{d\lambda}{db}=b\frac{d\lambda/dz}{db/dz},
\quad \lambda(z)=e^{\phi(z)},
\label{eq:betaf}
\ee
with the source term for $\lambda$ scaling like $N_c g^2$. Finally, the dynamics of the model is required to reproduce the logarithmic running of the coupling at $z\rightarrow 0$,
\be
\lambda(z)=\frac{1}{b_0\ln(1/(\Lambda z))}+\dots,
\ee
where one can choose to work either to one- or two-loop order. This determines the unit of energy $\Lambda$, which for the remainder of this section we choose as $\Lambda=1$.

The single most important quantity determining the dynamics of the model is its potential $V(\lambda)$. Its behavior at small $z$ is dictated by the running of the coupling through Eq.~\eqref{eq:betaf}, while in the far infrared, i.e.~for large $z$, the behavior of $V(\phi)$ is constrained by requiring confinement \cite{Gursoy:2007er}. The potential we employ reads \cite{Alho:2015zua}
\be
V(\lambda)= 12 \left( 1 + \dfrac{88 \lambda}{27} + \dfrac{4619 \lambda^2}{729 (1+ 2\lambda)} + 3 e^{-1/2\lambda} (2\lambda)^{4/3} \sqrt{1 + \log (1+2\lambda)} \right),
\label{eq:dilatonpot}
\ee
which in addition to satisfying the infrared and ultraviolet constraints has been fitted to lattice data for large-$N_c$ pure YM
theory~\cite{Panero:2009tv}.

As we shall demonstrate in the next section, the different parameters in the potential have been chosen so that the order of the phase transition and the equation of state are reproduced to sufficient accuracy. Varying them allows one to describe phase diagrams with higher order or even crossover transitions~\cite{Alanen:2009xs}. More generally, tuning the functional form of the dilaton potential $V(\lambda)$ provides holographic realizations of theories with different types of renormalization group flow both in the infrared~\cite{Alanen:2009na, Alanen:2011hh} and ultraviolet~\cite{Rey:2019wve} (see also the discussion in \cite{Gursoy:2010jh}).

As a final remark, we note that the model can be extended by adding another scalar field sourcing the $\bar{q}q$ operator and allowing for a full
treatment of fermion backreaction, resulting in a model commonly referred to as V-QCD \cite{Jarvinen:2011qe}. The phase diagram of this extended model has been studied at both at finite temperature~\cite{Alho:2012mh} and density~\cite{Alho:2013hsa}, and it has been extensively used in particular in the description of the dense nuclear and quark matter found inside compact stars \cite{Jokela:2018ers,Ishii:2019gta,Chesler:2019osn}.

\subsection{Numerical solution for thermodynamics}
\label{sec:numericalthermo}

To find a numerical solution to Eqs.~\eqref{Eq:Einstein}, it is useful to first define the new variable
\be
W=-\frac{\dot{b}}{b^2},
\ee
so that the equations are transformed to the first order forms
\begin{align}
	\dot{W} &= 4bW^2-\frac{1}{f}(W\dot{f}+\frac{1}{3}bV),\\
	\dot{b} &= -b^2W, \\
	\dot{\lambda} &= \frac{3}{2}\lambda\sqrt{b\dot{W}},\\
	\ddot{f} &= 3\dot{f}bW
\end{align}
that we can approach in a relatively straightforward manner (see e.g.~\cite{Alanen:2011hh} for details). The horizon values of the fields $b(z)$, $f(z)$ and $W(z)$, with the horizon located at $z=z_h$, are then obtained requiring finiteness and that the leading dependence of $\lambda$ on the energy scale in the UV be equivalent with the known behavior of the YM gauge coupling at two-loop order.

After generating a family of solutions parametrised by the values $\lambda_h\equiv \lambda(z_h)$, the thermodynamic behavior of the model can be determined from the relations
\begin{align}
	4\pi T &= -\dot{f}(z_h),\,\, s=\frac{b^3(z_h)}{4G_5},\label{eq:thermobasict} \\
	p &= \frac{1}{4G_5}\int_{\lambda_h}^\infty d\lambda_h'\left(-\frac{dT}{d\lambda_h'}\right)b^3(\lambda_h'),\label{eq:thermobasicp}\\
	\epsilon &= Ts-p.\label{eq:thermobasiceps}
\end{align}
The overall scale of thermodynamic quantities is affected by the choice of $4G_5$ in the above equations. In principle, this choice should be made by  matching with the noninteracting Bose-Einstein limit at asymptotically high temperatures, but to follow standard conventions in the field, we instead choose the scale by optimizing the model's fit to lattice data at temperatures only slightly above the critical temperature $T_c$ of the first-order deconfinement transition. This leads to overshooting the expected high-$T$ behavior of thermodynamic quantities by ca.~30\%, but as we will see in the following section, produces lower-temperature thermodynamics in excellent agreement with lattice predictions.

\subsection{Quasinormal modes}

Being equipped with the solved gravity background as well as with the potential fitted to lattice thermodynamics, we can now proceed to study the QNM spectra predicted by IHQCD. Here, we specialize to a field theory operator dual to a gauge-invariant scalar fluctuation $\phi(\omega,z)$, noting that at zero momentum and in the absence of mixing with
the metric fluctuations
the equation of motion for this operator takes the form
\be
\ddot{\phi}+\frac{d}{dz}\ln(fb^3)\dot{\phi}+\frac{\omega^2}{f^2}\phi=0.
\label{eq:fluct}
\ee
To aid the forthcoming analysis, we transform this equation into a Schr\"odinger-type form, with $\omega^2$ playing the role of an eigenvalue. This is
achieved by introducing the new variable
\be
u=\int_0^z\frac{dz'}{f(z')}
\ee
and defining $\psi(u)=\sqrt{b^3}\phi(u)$, whereby the equation of motion becomes
\begin{eqnarray}
&&-\psi''(u)+V_{\rm{Sch}}(u;z_h)\psi(u)=\omega^2\psi(u),\\
&&V_{\rm{Sch}}(u,z_h)=f^2\left[\frac{2\ddot{b}}{2b}+\frac{3\dot{b}^2}{4b^2}+
\frac{3\dot{f}\dot{b}}{2fb}\right]_{z=z(u)}.
\label{eq:effshrode}
\end{eqnarray}

Solving for the QNMs from this equation is in principle straightforward: the equation is linear, and its solution (for a fixed $\omega$) therefore fully determined by two complex numbers. One of them corresponds to the normalization, while the other can be determined by requiring the solution to be ingoing at the horizon, i.e.~ensuring that $\psi$ be proportional to $e^{+i\omega u}$ at large $u$. To obtain the QNMs, we then simply need to determine those values of $\omega$, for which the solutions satisfying this boundary condition are normalizable when $u\rightarrow 0$. To implement this boundary condition in the numerical calculation
it is convenient to find the solution in the infrared analytically and
match it with the numerically determined ultraviolet solution.
The details required in implementing the boundary condition are given in Appendix \ref{sec:app}.

A practical complication in the computation arises from the fact that standard spectral methods rely on the ability to expand solutions around $u\approx z=0$ in power series, while in the gravity model we are working with these expansions also contain logarithmic terms. This issue can, however, be resolved through the application of more robust numerical methods to remove the non-normalizable component from the solution. Here, it is crucial to recall that in a numerical calculation the solutions always contain a small part of the non-normalizable solution. However, if the solution is close to the correct one, the divergence due to the non-normalizable solution appears only when $u$ is very small. Then, the method introduced in~\cite{Alanen:2011hh} for finding the QNMs proceeds as follows: for a trial value of $\omega$, find the numerical solution towards the boundary and determine the minimum of $|\psi(u)|$, denoting its location by $u_{\rm{min}}$. The desired QNM is then approximately the value which minimizes $u_{\rm{min}}(\omega)$. For a detailed explanation of this
method, see Appendix B of \cite{Alanen:2011hh}.

\section{Results}
\label{sec:results}

\subsection{Thermodynamic properties}

We begin our discussion from thermodynamic quantities, which we however cover only rather briefly, as they have been extensively
considered in the literature (see e.g.~\cite{Alho:2012mh}). Indeed, our focus is mainly on those numerical results that serve as important consistency checks for our model and establish a connection between its finite-temperature phase structure and its spectrum as a function of $T$.

\begin{figure}[t!]
	\begin{center}
		\includegraphics[width=.45\textwidth]{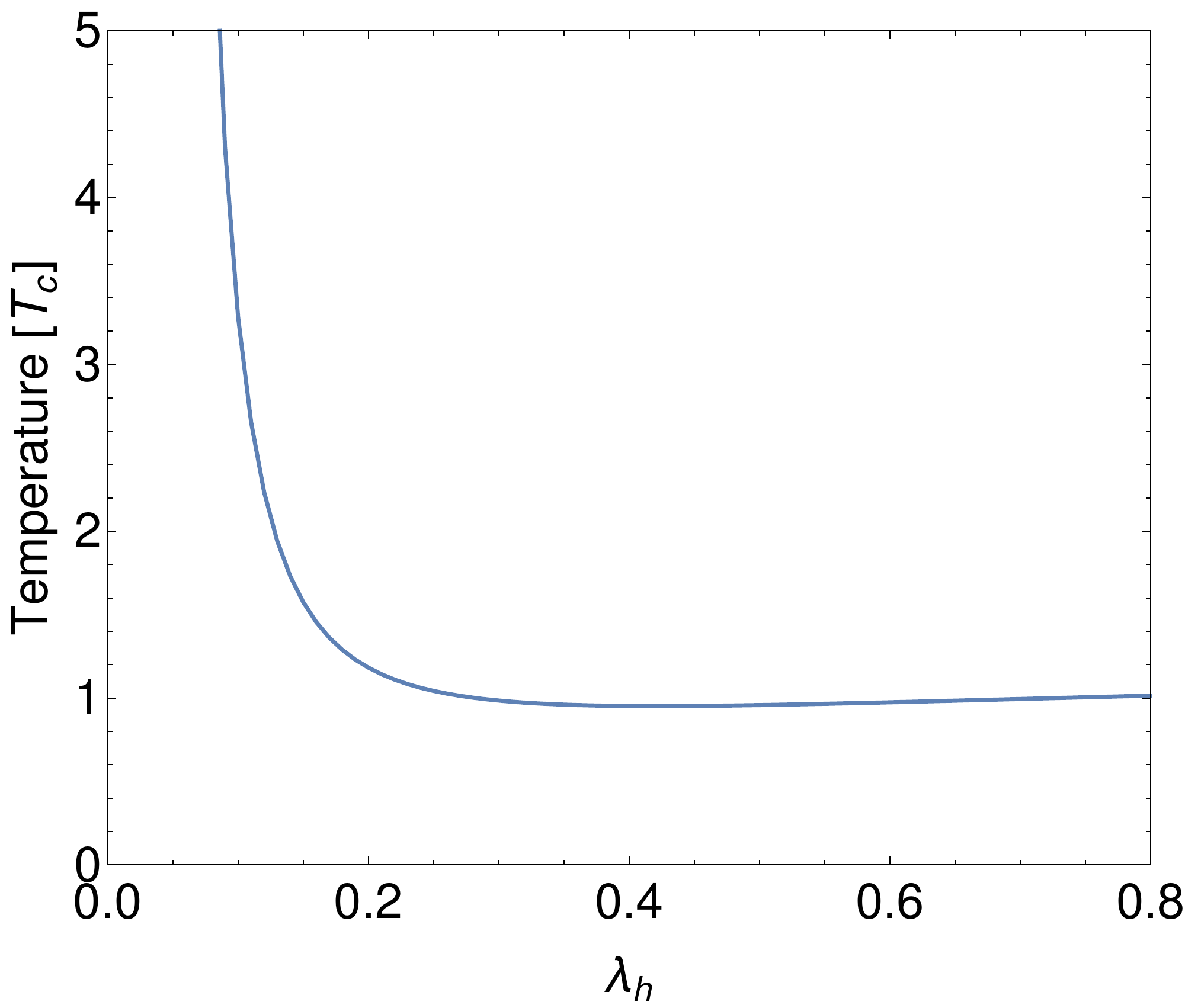}
		\quad
\includegraphics[width=0.45\textwidth]{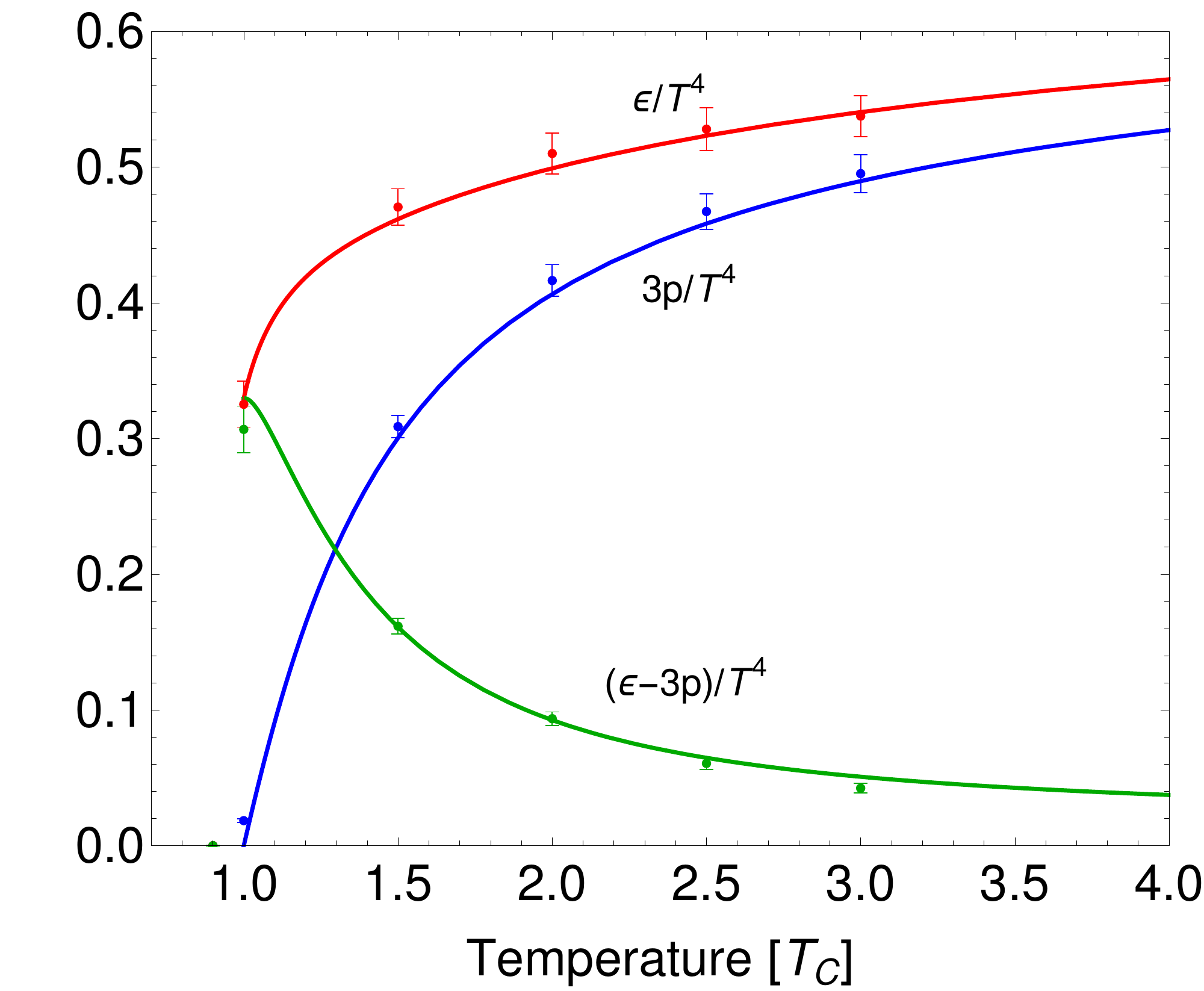}
		\caption{Left:  The temperature $T$ plotted as a function of $\lambda_h$, showing both the stable and unstable solutions to the left and right of the minimum, respectively. Right: Thermodynamic quantities as obtained from our model. The different curves correspond to the pressure $p$, energy density $\epsilon$, and the trace of the energy-momentum tensor $\epsilon-3p$ as functions of $T/T_c$, all
		normalized by $T^4$. Also shown is lattice data from Ref.~\cite{Panero:2009tv} with only the statistical error bars indicated. \label{fig:thermodynamics}
		}
	\end{center}

\end{figure}

Specifically, we want to compute the free energy, i.e.~the pressure $p(T)$, as defined in Eqs.~\eqref{eq:thermobasicp} in terms of the functions $b(\lambda_h)$ and $T(\lambda_h)$. Here, $b(\lambda_h)$ is a monotonous function, while $T(\lambda_h)$ first decreases with increasing $\lambda_h$ but then starts to increase; see the illustration in Fig.~\ref{fig:thermodynamics} (left). The behavior in the UV, or small $\lambda_h$, corresponds to the weak-coupling limit, i.e.~to large temperatures, while the domain towards the IR, where temperature increases with $\lambda_h$, is unstable. This can be seen e.g.~by computing the speed of sound squared: it turns out that this quantity is proportional to $-T^\prime(\lambda_h)$, so we must clearly require $T^\prime(\lambda_h)<0$.

Evaluating the pressure from high to low temperatures, we next determine the critical temperature as the point where the pressure becomes negative. From this condition, we find that
\be
T_c=0.7546\Lambda,
\ee
which allows us to express all dimensionful quantities in units of $T_c$.

Our results for the thermodynamics of the model are displayed and compared to the lattice data of \cite{Panero:2009tv} in Fig.~\ref{fig:thermodynamics} (right). From here, we observe that our results for the three key quantities --- the pressure $p$, energy density $\epsilon$ and trace anomaly $\epsilon -3p$ --- are in very good agreement with the lattice calculation. In particular, the slopes of all three functions are  accurately reproduced at temperatures somewhat above the critical one, which is a strong indication that our choice for the dilaton potential in Eq.~\ref{eq:dilatonpot} was indeed a reasonable one. As mentioned in Sec.~\ref{sec:numericalthermo}, the price to pay for treating the overall normalization of the pressure as a free parameter is that our results eventually overshoot the expected high-temperature limits of these quantities by some 30\%. At the same time, it interestingly turns out that with our normalization convention the UV behavior of energy momentum tensor correlators becomes very accurately reproduced  \cite{Kajantie:2013gab}.

\begin{figure}[t!]
\begin{center}
\includegraphics[width=0.45\textwidth]{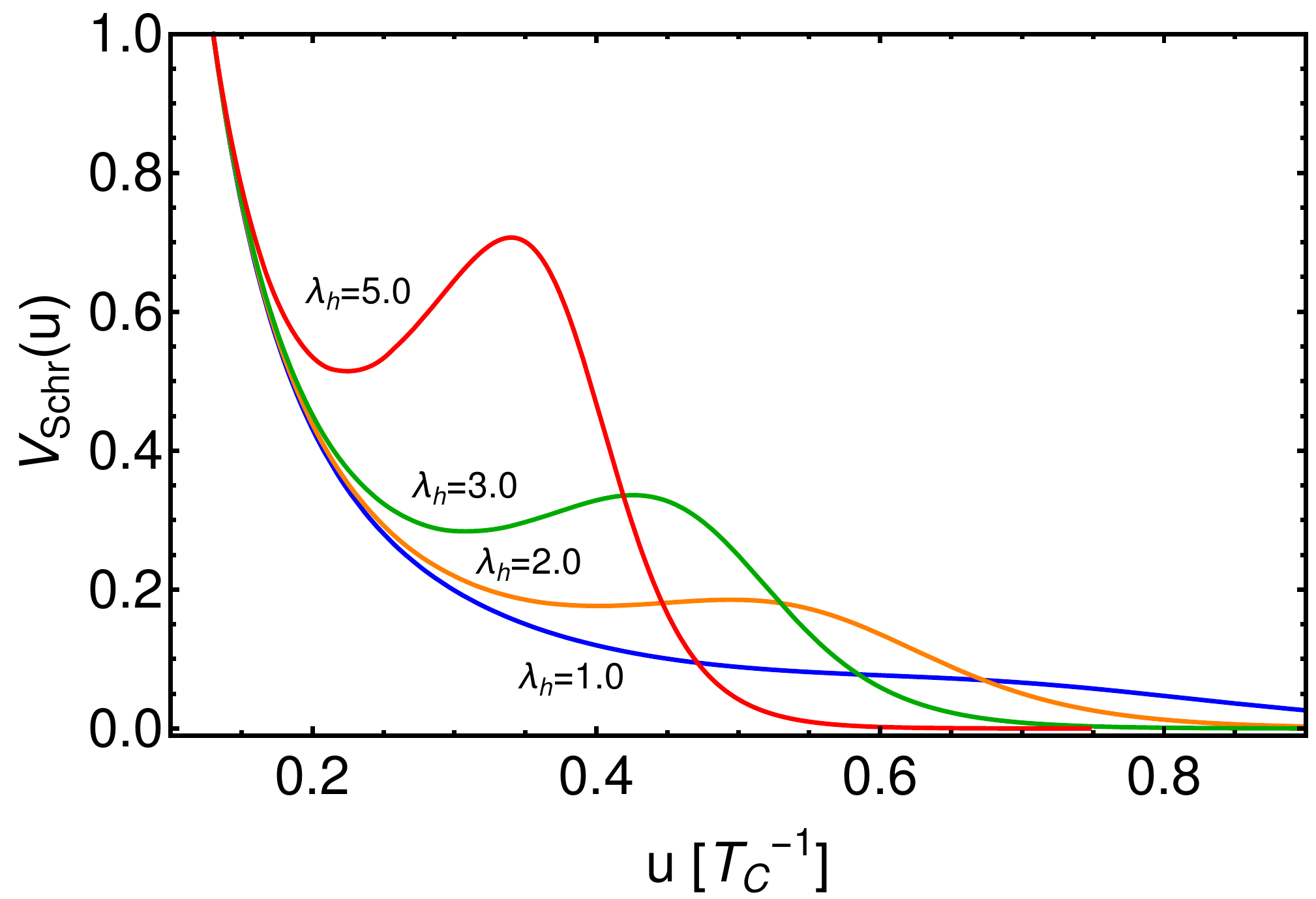}
\quad
\includegraphics[width=0.45\textwidth]{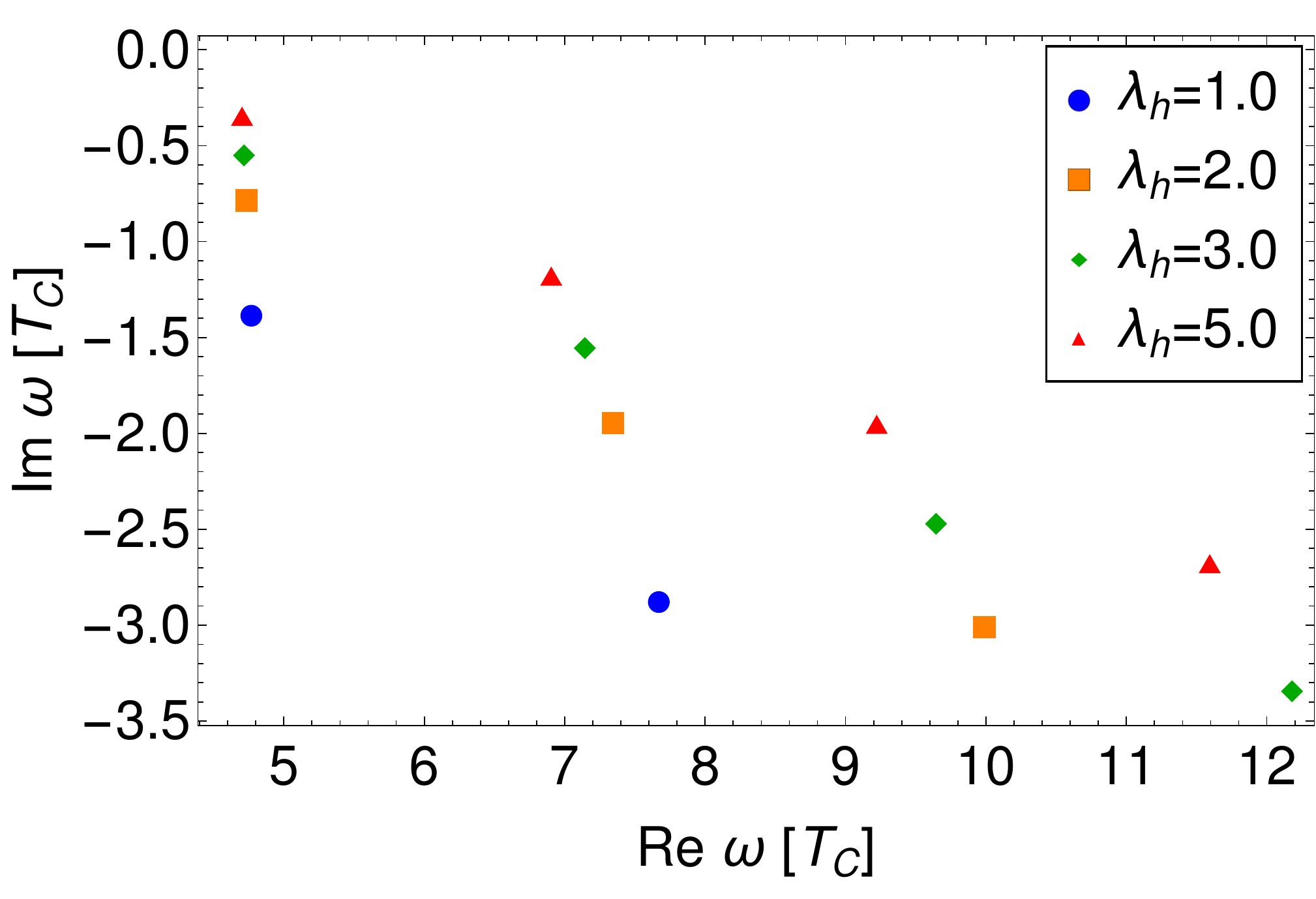}
\end{center}
\caption{Left: Schr\"odinger potentials corresponding to the high-$\lambda_h$ unstable branch and normalized at $u=0.1$ for illustrative purposes. Right: the corresponding QNMs, displayed using the same color coding  that was introduced for the potentials.}
\label{fig:highlah}
\end{figure}

\subsection{Spectrum of quasinormal modes}

Moving next on to the QNM spectra, we first note that at zero temperature, corresponding to $\lambda_h\rightarrow\infty$ and $f=1$, the potential entering the Schr\"odinger equation is well approximated by (see~\cite{Alanen:2011hh})
\be
V_{\rm{Sch}}(u;z_h)\simeq \frac{15}{4u^2}+2+u^2,
\ee
from which one obtains a bound state spectrum with masses
\be
m_n=2\sqrt{2+n},\quad n=0,1,2,\dots
\ee
As $\lambda_h$ is lowered, a numerical calculation performed on the unstable branch of the theory
shows how the potential is perturbed away from the harmonic-oscillator-like form as depicted in the left panel of Fig.~\ref{fig:highlah} (note, however, that the instability analyzed in \cite{Janik:2016btb,Gursoy:2016ggq} is not present for the potentials we consider). The corresponding spectrum of QNMs,  shown in the right panel of Fig.~\ref{fig:highlah}, displays a clear and expected pattern: the spectrum moves away from the real axis, with the states becoming broader. It should be highlighted, however, that these results,
derived on the unstable branch, can not be straightforwardly related to the physics of the YM theory. Nevertheless, these results connect smoothly with the
quasinormal modes determined on the high temperature branch where the system enters the stable high-temperature phase. There the states are observed to
become even broader, eventually melting away.

\begin{figure}[t]
\begin{center}
\includegraphics[width=0.45\textwidth]{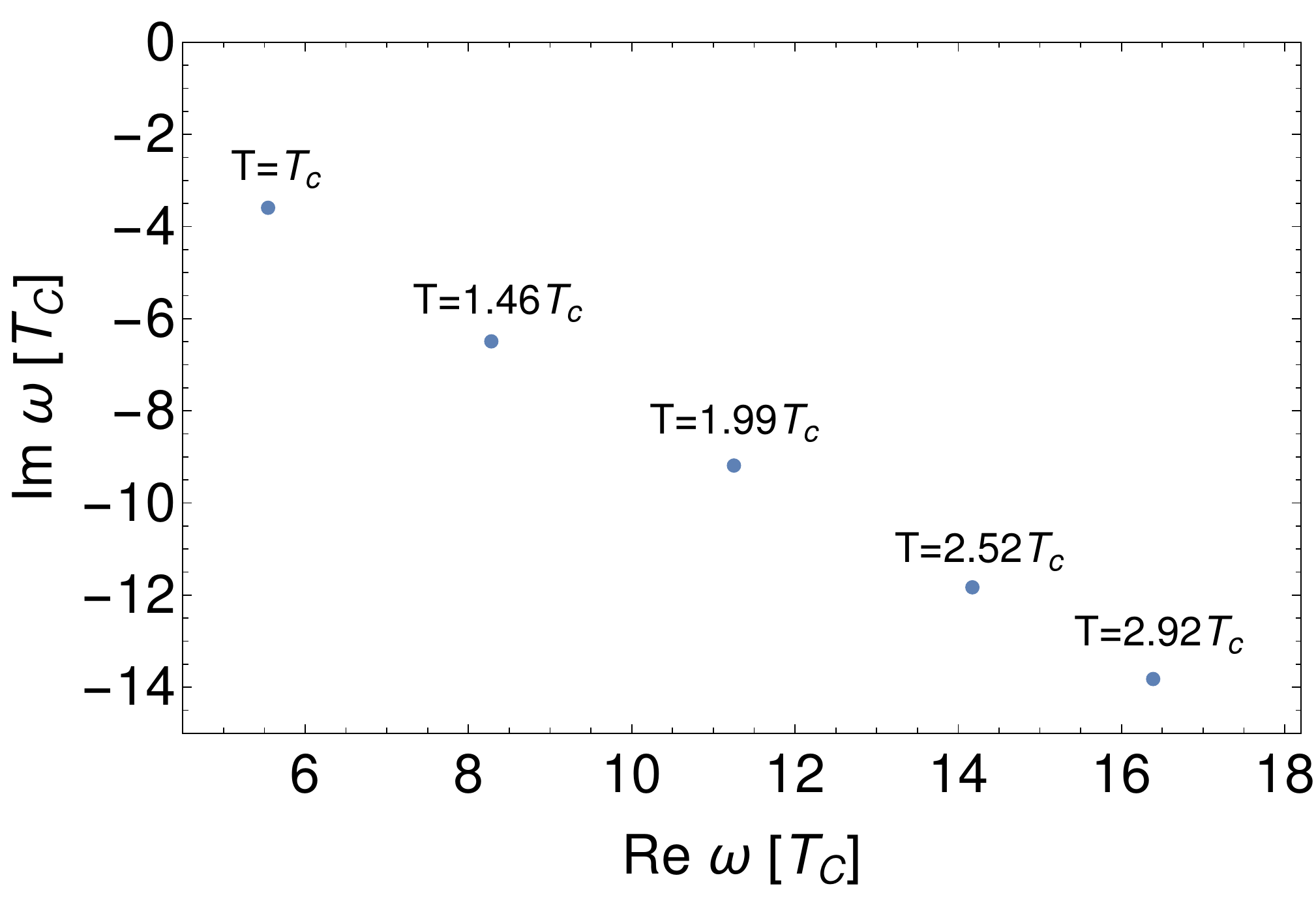}
\quad
\includegraphics[width=0.45\textwidth]{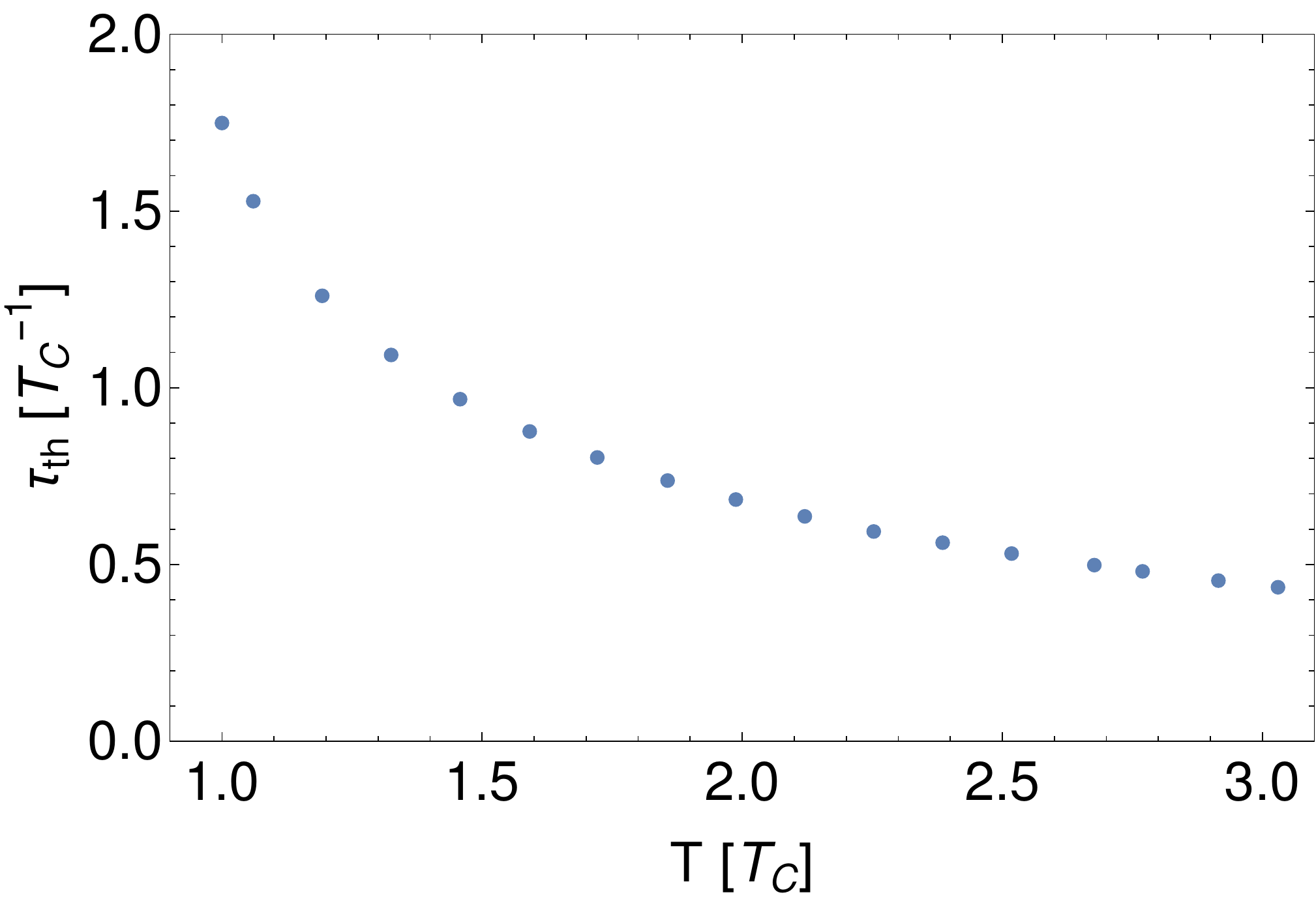}
\end{center}
\caption{Left: location of the lowest quasinormal mode $\omega$ shown for five different temperatures on the stable BH branch. Right: thermalization time $\tau_{th}$ in units of $T_c^{-1}$ versus the temperature in units of $T_c$.}
\label{fig:therm}
\end{figure}

As our physical interest lies in the description of thermalization dynamics at temperatures somewhat (but not excessively) above the critical one, we next specialize to the stable branch of the theory and analyze in more detail the QNM spectrum for temperatures $T_c<T< 3 T_c$. Of particular interest here is the lowest QNM and specifically its imaginary part, which has been argued to be inversely proportional to the equilibration time for the correlator in question \cite{Keranen:2015mqc}. For temperatures ranging from $T_c$ to $3T_c$, we first display the lowest QNMs on the complex-$\omega$ plane in the left panel of Fig.~\ref{fig:therm}, observing that the lowest mode roughly follows the trend ${\rm{Im}}\,\omega\approx -i{\rm{Re}}\,\omega$ and $\omega\approx 2\pi T(1-i)$. The thermalization time obtained from the lowest mode via the relation \cite{Keranen:2015mqc}
\be
\tau_{\rm{th}}= -\frac{2\pi}{{\rm{Im}}\,\omega}
\ee
is shown as a function of temperature in the right panel of Fig.~\ref{fig:therm}. It is observed to decay slightly faster than $1/T$, with the scale being of order  $\tau_{\rm{th}}\simeq 0.5/T_c\approx 0.5$
fm/c, which is a phenomenologically very reasonable result.

Finally, an interesting observation can be made about the fate of higher QNMs on the stable BH branch as a function of increasing temperature. In our numerical calculation, we find that the string of clearly separated, individual QNMs terminates at a structure that one is tempted to interpret as an emerging branch cut on the complex $\omega$-plane; cf.~Fig.~\ref{fig:therm2}. This structure always corresponds to a constant value of ${\rm{Im}}\,\omega$, with the number of accessible separate QNMs decreasing with increasing $T$.
This behavior is consistent with the existence of quasistable states as obtained from Eq.~\eqref{eq:effshrode}, with the potential evolving as shown in Fig.~(\ref{fig:highlah}) (for the unstable branch). We have checked that the
existence of this structure is independent of whether we use a leading or next-
to-leading order expansion for the analytic solution of the Schr\"odinger
equation in the infrared.

\section{Conclusions}
\label{sec:checkout}

In the holographic study of gauge theory equilibration, one can identify two somewhat distinct long-term goals. On one hand, there have been frequent attempts to make the physical setting under study  more closely resemble
realistic heavy ion collisions by considering the collision of shock waves of finite thickness, transverse size, and anisotropy \cite{Chesler:2010bi,Casalderrey-Solana:2013aba,Bantilan:2017kok,Waeber:2019nqd}. At the same time, many groups have been working on modifying the gravity background in the shock-wave setting to have the dual gauge theory feature broken conformality or supersymmetry, or even non-infinite coupling strengths \cite{Grozdanov:2016zjj,Attems:2016ugt,Attems:2016tby,Attems:2017zam,Folkestad:2019lam}. What is, however, common to both of these lines of work is that one usually works in firmly top-down settings, typically starting from the conformal and supersymmetric ${\mathcal N}=4$ SYM theory and departing from it one step at a time, which makes progress towards the description of a truly QCD-like theory rather slow.

\begin{figure}[t!]
\begin{center}
\includegraphics[width=0.3\textwidth]{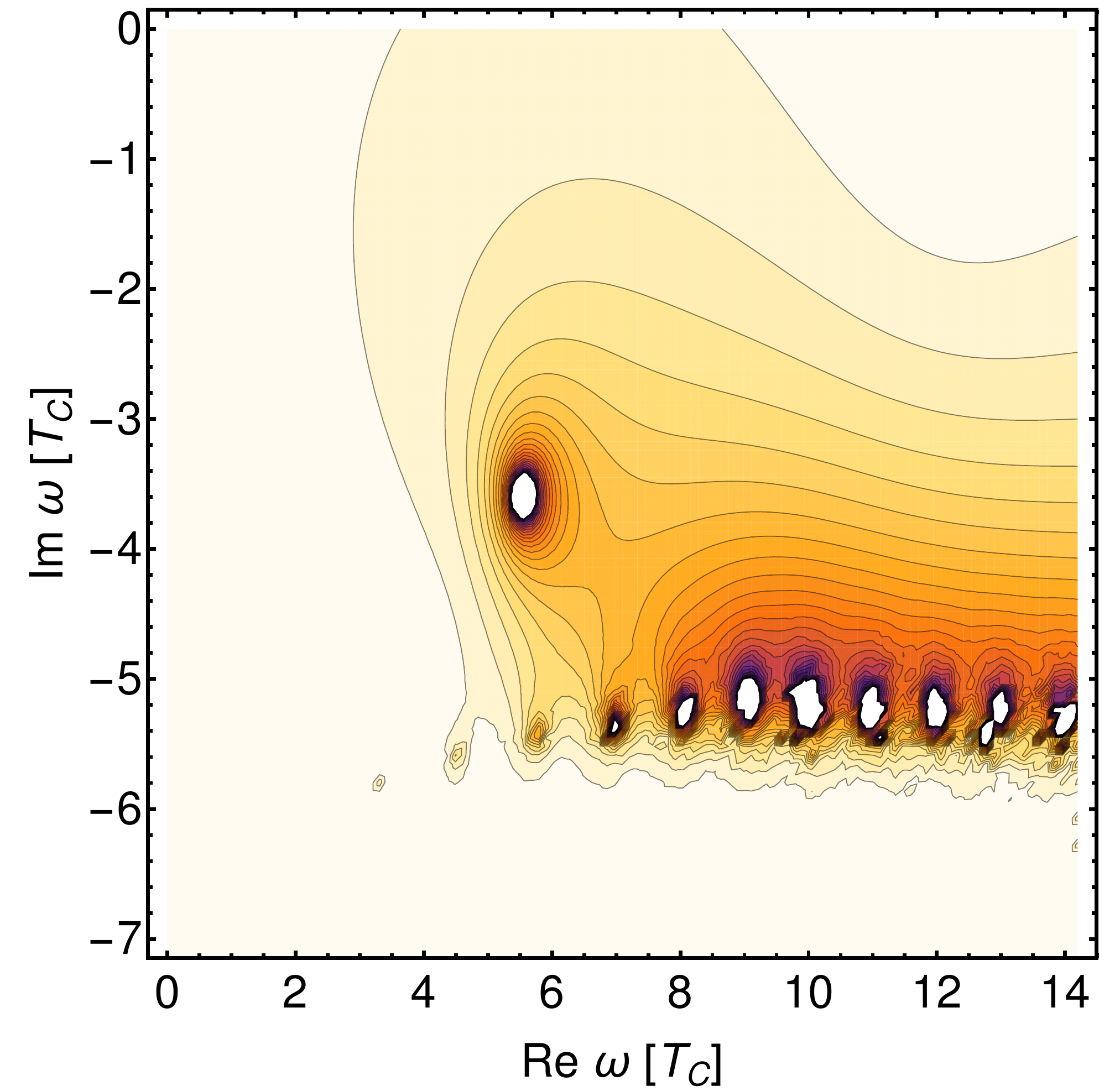}
\quad
\includegraphics[width=0.3\textwidth]{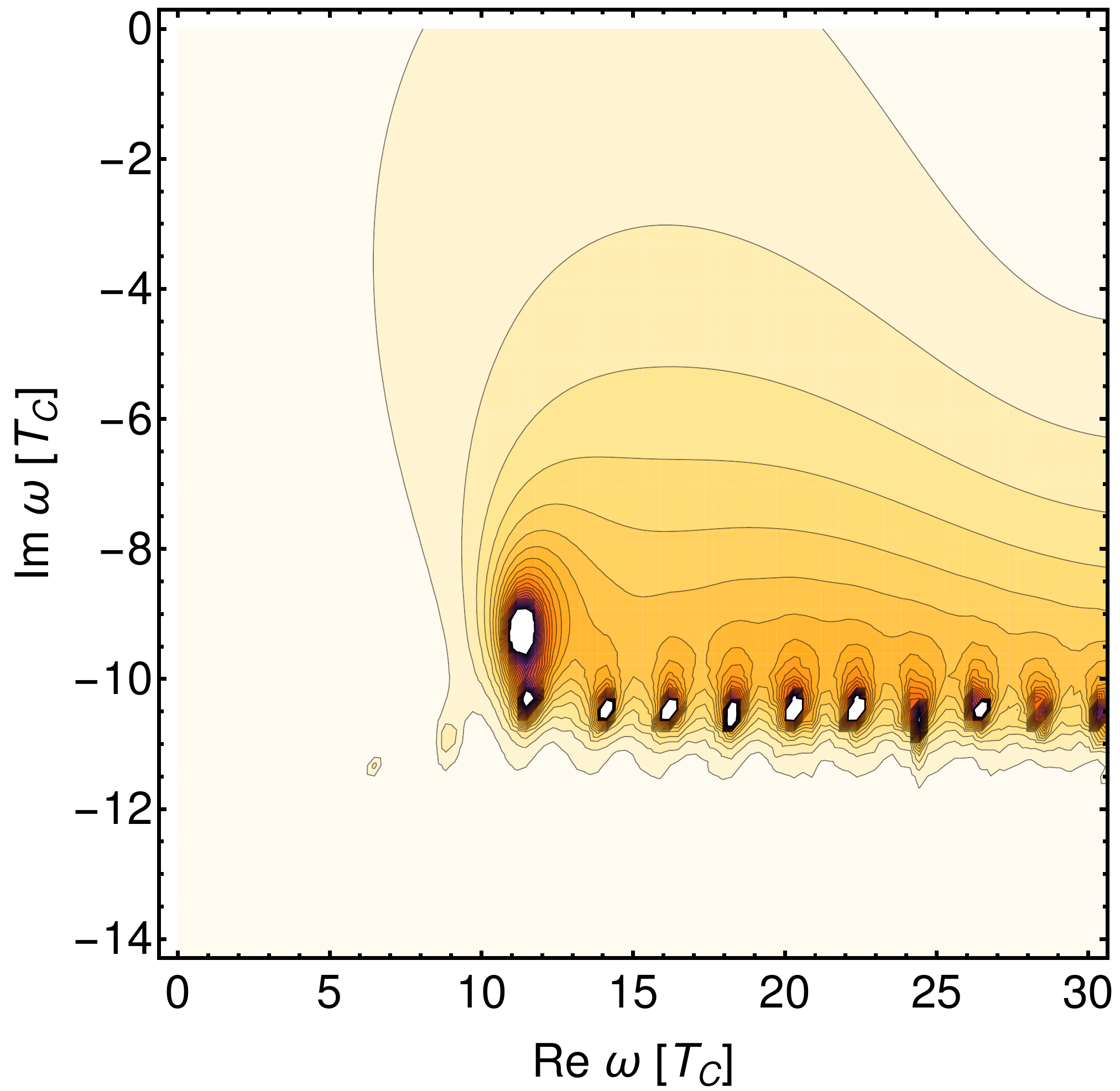}
\quad
\includegraphics[width=0.3\textwidth]{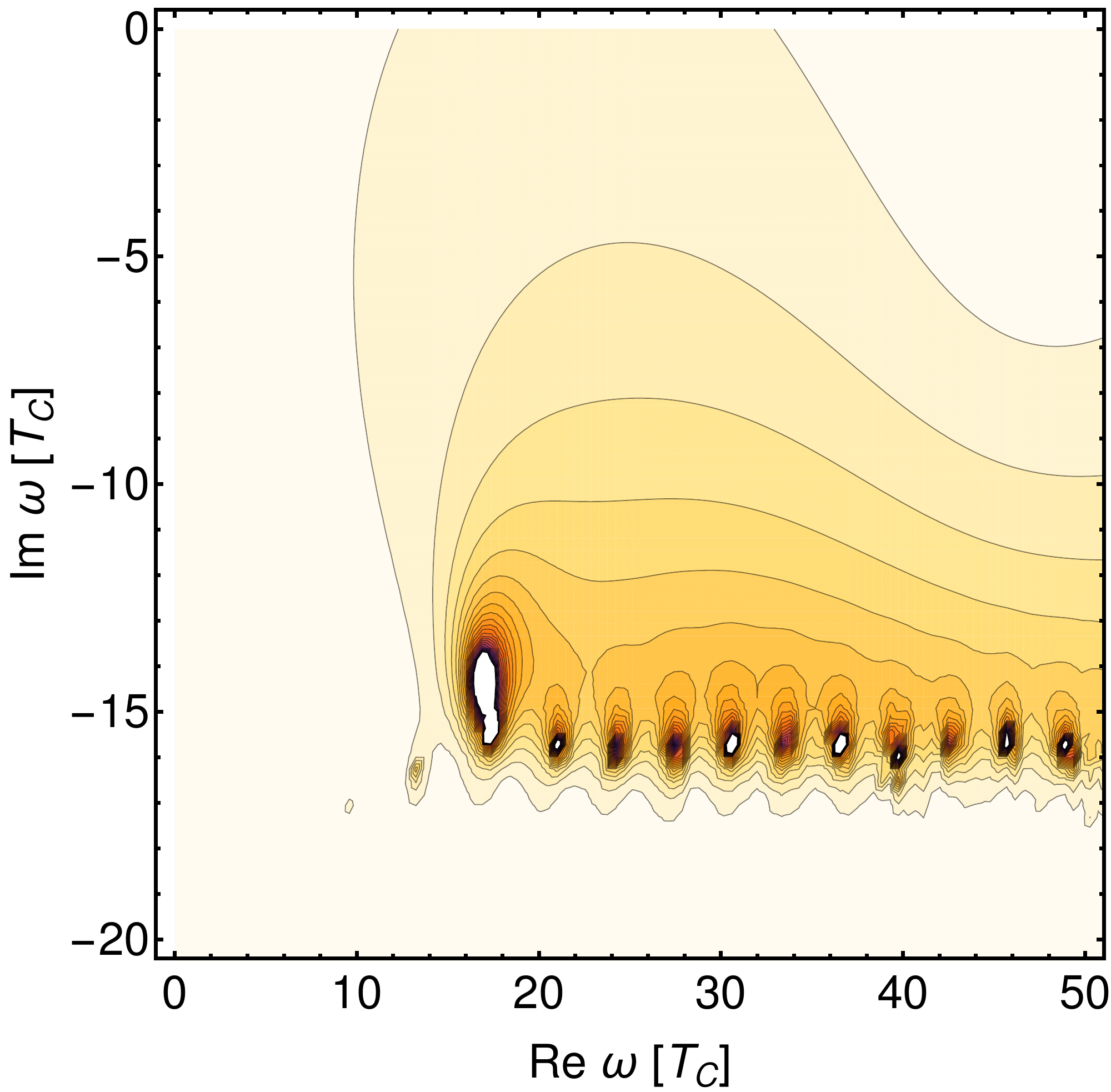}
\end{center}
\caption{Contour plots of $u_{min}^{-1}$ on the stable BH branch on the complex $\omega$-plane at $T=T_c$, $2T_c$, and $3T_c$ (from left to right).}
\label{fig:therm2}
\end{figure}

Independently from the study of shock-wave collisions and other developments in applied top-down holography, systematic efforts towards constructing a bottom-up model to closely resemble non-supersymmetric Yang-Mills theory and QCD have culminated in the development of the Improved Holographic QCD and V-QCD frameworks that have been demostrated to faithfully reproduce the equilibrium thermodynamic properties of the full theories in the large-$N_c$ limit \cite{Gursoy:2007cb,Gursoy:2007er,Alho:2012mh,Alho:2013hsa}. Considering the technical complexity of building a computational framework for studying shock-wave collisions in this type of a setting, it is clearly worthwhile to investigate equilibration within the IHQCD model in a somewhat more modest way: through the study of quasinormal modes that describe the response of the dual field theory system to small departures from equilibrium.

In the paper at hand, we have determined the QNM spectrum for one particular gauge theory operator in the IHQCD framework, dual to a scalar field on the gravity side, for a range of temperatures from $T_c$ up to ca.~$3T_c$. In particular, we have studied the temperature dependence of the lowest nonzero QNM to obtain an estimate for the thermalisation time of the corresponding correlator, finding phenomenologically very reasonable results. In addition, we have observed that the number of clearly separate QNMs decreases with increasing temperature --- a result we have been able to link to the broadening and melting of the corresponding states. Instead of a clean spectrum of higher individual QNMs separated from each other by the expected $2\pi T$, we have witnessed the emergence of a linear structure parallel to the real axis of the complex frequency plane, which may point towards to the presence of a brach cut. In this context, it is worth pointing out that while we have only shown results for the stable BH branch in Fig.~\ref{fig:therm2}, the same qualitative conclusions apply to the unstable branch as well, with the main difference being the somewhat higher number of individual QNMs there.

It is clearly very interesting to contrast our findings with the results of similar exercises carried out in the past. Our present work can be regarded as a rather direct continuation of a series of works by other groups, in which somewhat simpler versions of IHQCD (or other closely related models) have been considered \cite{Alanen:2011hh,Janik:2016btb,Betzios:2017dol,Betzios:2018kwn}, although it should in addition be recalled that  some of these works considered correlators different from the one studied by us. Of these papers, Alanen et al.~\cite{Alanen:2011hh} and Janik et al.~\cite{Janik:2016btb} both employed the IHQCD framework but with simpler choices of the potential, with e.g.~\cite{Janik:2016btb} not enforcing the condition that the gauge coupling should run logarithmically in the UV. Neither of these references reported the existence of a branch-cut like structure, which is likely related to the IR behaviors of the employed potentials. In this respect, a very interesting point of comparison is the work of Betzios et al.~\cite{Betzios:2017dol}, which considered a different model of Einstein-dilaton gravity, dual to a Chamblin-Reall plasma in the IR and having the usual AdS form in the UV. Similarly to us, they witnessed the  emergence of a branch cut in the critical limit, albeit exactly on the real axis. It would clearly be very interesting to analyze the reason for this qualitative similarity in a more explicit manner, but we leave this for future work.

Finally, we note that while in the calculation presented above we have worked in the context of pure gauge theory, our results can be extended to QCD in the Veneziano limit by employing the V-QCD model \cite{Alho:2013hsa}. Performing a similar study of QNM spectra in this framework and thereby analyzing the effect of dynamical quarks on equilibration would clearly be a very interesting, albeit technically more demanding exercise to carry out.

\section*{Acknowledgments}

We thank Niko Jokela and Matti J\"arvinen for enlightening discussions and useful comments on the manuscript. The work has been supported by the European Research Council, grant no.~725369, and by the Academy of Finland grants no.~1322507 and 1310310.

\appendix

\section{Asymptotic solutions}
\label{sec:app}

In this Appendix, we briefly discuss the asymptotic expansions of the bulk fields, which  turned out to be very useful in the numerical determination of the QNM spectra. In this context, we note that when dealing with the Schr\"odinger-type equation (\ref{eq:effshrode}), it is useful to first expand the potential at $u \to \infty$ and solve the equation analytically in this limit. However, this is done more easily after introducing a new variable $A$ such that $b(z) = \exp (A(z))$ and solving the equation near the limit $A \to A_h$. For this purpose, we also need to introduce the function
 \begin{equation}
 q(A) = e^{A} \frac{dz}{dA} ,
 \end{equation}
 so that the fluctuation equation~\eqref{eq:effshrode} obtains the form
 \begin{equation}
 \psi '' (A) + \left( 4 + \dfrac{f'(A)}{f(A)} - \dfrac{q'(A)}{q(A)} \right) \psi'(A) + \dfrac{e^{-2A} q(A)^2}{f(A)^2} \omega^2 \psi(A) = 0.
 \end{equation}

 To proceed from here, we note the relation
 \begin{equation}
  u(A) = \int_{\infty}^{A} e^{-\tilde{A}} \frac{q(\tilde{A})}{f(\tilde{A})} \dif{\tilde{A}} = \int_{\infty}^{A} \frac{1}{\tilde{A}} e^{-\tilde{A}} \frac{q(\tilde{A})}{\hat{f}(\tilde{A})} \dif{\tilde{A}},
 \end{equation}
 where $\hat{f}(A) \equiv \frac{f(A)}{A}$ and we have set $A_h = 0$ for convenience.
 Observing that $\hat{f}$ is regular and non-zero at $A = A_h = 0$, we can isolate the divergence by integrating by parts
 \begin{equation}
 \begin{split}
 u(A) =& \left.\left(\log(\tilde{A}) e^{-\tilde{A}} \frac{q(\tilde{A})}{\hat{f}(\tilde{A})}\right)\right\vert_{\infty}^{A} -\int_{\infty}^{A} \log(\tilde{A}) \left(e^{-\tilde{A}} \frac{q(\tilde{A})}{\hat{f}(\tilde{A})}\right)' \dif{\tilde{A}}\\
 =& \log(A) e^{-A} \frac{q(A)}{\hat{f}(A)} -\int_{\infty}^{0} \log(\tilde{A}) \left(e^{-\tilde{A}} \frac{q(\tilde{A})}{\hat{f}(\tilde{A})}\right)' \dif{\tilde{A}}
 \\
 &- \int_{0}^{A} \log(\tilde{A}) \left(e^{-\tilde{A}} \frac{q(\tilde{A})}{\hat{f}(\tilde{A})}\right)' \dif{\tilde{A}},
 \end{split}
 \end{equation}
 where the prime denotes differentiation with respect to $A$ and we have used the fact that $q$ and $\hat{f}$ grow more slowly  than the inverse of $\log(A) e^{-A}$ at large $A$.

 Next, we define
 \begin{equation}
  u_0 =  -\int_{\infty}^{0} \log(\tilde{A}) \left(e^{-\tilde{A}} \frac{q(\tilde{A})}{\hat{f}(\tilde{A})}\right)' \dif{\tilde{A}}
 \end{equation}
 and write
 \begin{equation}
 h(A) = e^{-A} \frac{q(A)}{\hat{f}(A)} = \sum_{k = 0}^\infty h_k A^k,
 \end{equation}
 where $h_k = \frac{1}{k!}h^{(k)}(A)\vert_{A = 0}$, which can be explicitly computed in terms of the near-horizon expansions of $q$ and $f$.
 Then,  we expand around $A = 0$ to get
 \begin{eqnarray}
 u(A) - u_0 &=& \log(A) h_0 + \sum_{k = 1}^\infty \left(\log(A) h_k A^k - \int_0^A \log(\tilde{A}) h_k k \tilde{A}^{k-1} \dif{\tilde{A}} \right) \nonumber\\
 &=& h_0 \log(A) + \sum_{k = 1}^\infty \frac{A^k}{k} h_k.
 \label{eq:horizonuser} \\
 &=& h_0 \log(A) + \int_0^A \frac{1}{\tilde{A}} (h(\tilde{A}) - h(0))\dif{\tilde{A}}. \label{eq:horizonuint}
 \end{eqnarray}
 Note that if $h(A)$ is real-analytic, the logarithm appears only in the leading term of the expansion.

 In order to write the Schr\"odinger equation in a more useful form, we have to invert the above relation, i.e.~find the function $A(u)$. To this end, we  write $\hat{u} = u - u_0$ to simplify the notation. To leading order, we can immediately solve
 \begin{equation}
 A = e^{\frac{\hat{u}}{h_0}},
 \end{equation}
 which is very small for large $\hat{u}$, as $h_0 < 0$.

 To get the next-to-leading order term, we attempt to substitute the leading term to Eq.~\eqref{eq:horizonuser} and expand to  order $A$. The resulting error term is $h_1 e^{\frac{\hat{u}}{h_0}} + O(A^2)$, which leads to the ansatz
 \begin{equation}
 A(\hat{u}) = e^{\frac{\hat{u}}{h_0} + u_1 e^{\frac{\hat{u}}{h_0}}}
 \end{equation}
 for some number $u_1$.
 Using the fact that $e^{\hat{u}/h_0}$ is small, we can expand
 \begin{equation}
 e^{e^{\frac{\hat{u}}{h_0}}} = 1 + e^{\frac{\hat{u}}{h_0}} + O(e^{2\frac{\hat{u}}{h_0}}),
 \end{equation}
 where the last term is of order $A^2$, leading to the ansatz
 \begin{equation}
 A(\hat{u}) = e^{\frac{\hat{u}}{h_0}} + u_1 e^{2 \frac{\hat{u}}{h_0}} + O(A^3).
 \end{equation}
 Inserting this again to Eq.~\eqref{eq:horizonuser}, we get
 \begin{equation}
 u(A) - u_0 = u + (u_1 h_0 + h_1)e^{\frac{\hat{u}}{h_0}} + O(e^{2\frac{\hat{u}}{h_0}}),
 \end{equation}
 which implies
 \begin{equation}
 u_1 = -\frac{h_1}{h_0}.
 \end{equation}
 It is clear that  from here the series would continue further in powers of $e^{\hat{u}/h_0}$.

 The numerics provide us the Schr\"odinger potential in terms of $A$.
 Writing
 \begin{equation}
 V_\mathrm{Schr}(A) = V_\mathrm{Schr}'(0) A + \frac{1}{2}V_\mathrm{Schr}''(0) A^2 + O(A^3)
 \end{equation}
 and inserting the expansion above gives
 \begin{equation}
 V_\mathrm{Schr}(u) = V'_\mathrm{Schr}(0) e^{\frac{\hat{u}}{h_0}} + \left(V'_\mathrm{Schr}(0) u_1 + \frac{1}{2} V''_\mathrm{Schr}(0) \right)e^{2\frac{\hat{u}}{h_0}} + O(A^3) \label{eq:schruexpanse}.
 \end{equation}
 The corresponding equation can be solved analytically, giving
 \begin{equation}
 \begin{split}
 \psi(u) =
  &C_1 U\left(\alpha,\beta,\gamma\right)e^{i\zeta}
 +C_2
 L_{i\mu}^{i\nu }\left(\rho\right) e^{i\zeta}
 \end{split}
 \end{equation}
 where $U$ is Tricomi's confluent hypergeometric function, and $L$ is is the generalized Laguerre polynomial.
 We have defined the following shorthand notations:
 \begin{align*}
 \alpha &= \frac{i}{2 \sqrt{V_2}}
 \left(-i V_1 h_0+2 \omega  \sqrt{V_2} h_0-i \sqrt{V_2}\right) \\
 \beta &= 2 i \omega  h_0+1 \\
 \gamma &= 2 e^{\frac{\hat{u}}{h_0}}
 h_0 \sqrt{V_2}.\\
 \zeta &=  h_0 \left(\omega  \log \left(e^{\frac{\hat{u}}{h_0}}\right)+i \sqrt{V_2} e^{\frac{\hat{u}}{h_0}}\right)\\
 \mu &=i\alpha\\
 \nu &= 2 h_0 \omega\\
 \rho &= 2 h_0 \sqrt{V_2}
 e^{\frac{\hat{u}}{h_0}}.
 \end{align*}
 Furthermore, $V_1$ and $V_2$ are the coefficients appearing in \eqref{eq:schruexpanse} and $C_1$ and $C_2$ are constants of integration.

 Finally, we note for reference the relations
 \begin{align}
 h_0 &= \frac{q_h}{f'_h}\\
 h_1 &= -\frac{q_h + \frac{f''_h}{2 f'_h} - q'_h}{f'_h}
 \end{align}

\end{document}